\documentclass[12pt]{article}
\usepackage{amssymb}
\usepackage{graphicx}

\advance \topmargin by -\headheight
\advance \topmargin by -\headsep

\evensidemargin \oddsidemargin
\newcommand{\beq}{\begin{equation}}
\newcommand{\eeq}{\end{equation}}

\newcommand{\bear}{\begin{eqnarray}}
\newcommand{\eear}{\end{eqnarray}}

\def\ie{{\em i.e.}}
\def\ie{\hbox{\it i.e.}}

\def\CC{{\mathchoice
{\rm C\mkern-8mu\vrule height1.45ex depth-.05ex
width.05em\mkern9mu\kern-.05em}
{\rm C\mkern-8mu\vrule height1.45ex depth-.05ex
width.05em\mkern9mu\kern-.05em}
{\rm C\mkern-8mu\vrule height1ex depth-.07ex
width.035em\mkern9mu\kern-.035em}
{\rm C\mkern-8mu\vrule height.65ex depth-.1ex
width.025em\mkern8mu\kern-.025em}}}

\def\RR{{\rm I\kern-1.6pt {\rm R}}}

\def\ZZ{{\rm Z}\kern-3.8pt {\rm Z} \kern2pt}
\def\IB{\relax{\rm I\kern-.18em B}}
\def\ID{\relax{\rm I\kern-.18em D}}
\def\II{\relax{\rm I\kern-.18em I}}
\def\IP{\relax{\rm I\kern-.18em P}}

\def\prl{Phys. Rev. Lett.}

\def\mpl{Mod. Phys. Lett.}

\def\jhep{J. High Energy Phys.}

\def\jgp{J. Geom. Phys.}

\begin{document}

\begin{titlepage}

\begin{center} \Large \bf Adding flavor on the Higgs branch

\end{center}

\vskip 0.3truein
\begin{center}
Daniel Are\'an${}^{\,*}$\footnote{arean@fpaxp1.usc.es}

\vspace{0.3in}

${}^{\,*}$Departamento de F\'\i sica de Part\'\i culas, Universidade de
Santiago de Compostela \\and\\
Instituto Galego de F\'\i sica de Altas Enerx\'\i as (IGFAE)\\
E-15782 Santiago de Compostela, Spain

\end{center}
\vskip.8
truein

\begin{center}
\bf ABSTRACT
\end{center}
We study the holographic dual of the Higgs branch of
${\cal N}=4$ SYM in four dimensions coupled to quenched fundamental
matter. The fundamentals are added via flavor D7-branes and the Higgs
phase is obtained when the color D3-branes and the flavor D7-branes
recombine with each other.  The holographic dual consists of D7-brane
probes in the background generated by the color D3-branes. In this
supergravity dual the Higgs phase is realized by switching on a selfdual
field strength on the worldvolume of the  D7-branes. A complementary
microscopic description in terms of dielectric D3-branes is also
possible. This microscopic approach permits a better matching with the
dual field theory analysis. Finally, the meson mass spectrum is computed
by studying the fluctuations of the flavor D7-branes.

\vskip5.6truecm
\smallskip
\end{titlepage}
\setcounter{footnote}{0}

\section{Introduction}

In its original formulation, the AdS/CFT correspondence
\cite{dual} describes  gauge supersymmetric theories with all their fields
in the adjoint representation. A natural extension of the
duality consists of the inclusion of matter fields in the fundamental
representation, which corresponds to the addition of an open string
sector to the supergravity side.  This approach was initiated in
\cite{KR,KK,KKW} by considering the orthogonal intersection of $N_f$ Dq-branes
and $N$ Dp-branes with $N_f\ll N$. Once the decoupling limit is
taken, as $N_f\ll N$, the system will consist of $N_f$ Dq-brane
probes in the background given by the near horizon geometry of $N$
Dp-branes. The open string degrees of freedom living on the Dq-brane
probes are identified with the open strings stretching between the
Dp- and Dq-branes and thus correspond to fundamental hypermultiplets
added to the gauge theory on the worldvolume of the Dp-brane.
Moreover, if $q>p$ the decoupling limit forces the $SU(N_f)$ gauge
symmetry on the Dq-brane to decouple and this symmetry
becomes the global flavor symmetry of the fundamental
fields. In this context, the fluctuations of the flavor brane
(the Dq-brane) should correspond to mesons in the dual gauge theory.
This study was started in \cite{KMMW} for the D3-D7 intersection and it
was extended to different probe branes in several backgrounds (for a
review see \cite{RwAv, RwJ}).

The theories dual to the Dp-Dq intersections described above are at
the origin of the moduli space. However, one can consider more involved
situations such as Higgs phases. Here we will study the SUGRA dual of the
Higgs phase for the D3-D7 system. It was proposed in \cite{gural,EGG}
that one can realize a mixed Coulomb-Higgs branch by switching on an
instanton configuration of the worldvolume gauge field of the D7-brane.
We will give two complementary descriptions of the system: the
macroscopic description in terms of the flavor D7-branes and the
microscopic one in terms of dielectric D3-branes which become fuzzy along
their transverse directions and recombine into the flavor D7-brane. The
results presented here are part of a general analysis of the Dp-Dq
systems in the Higgs phase published in \cite{ARR} (see \cite{Rvdg} for a
review).

\section{D3-D7 intersection}
\label{d3d7inters}

In this section we will deal with the system given by the
intersection of $N$ D3-branes and $N_f$ D7-branes, with the D7s
being spacetime filling along the D3-branes. When $N_f\ll N$ the
decoupling limit of the system corresponds to having $N_f$ D7-brane
probes in the $AdS_5\times S^5$ geometry generated by D3-branes. It
can be seen that the dual gauge theory is ${\cal N}=4$, $d=4$ SYM
coupled to $N_f$ ${\cal N}=2$ fundamental hypermultiplets whose mass
is proportional to the separation between the D3- and D7-branes.

The classical SUSY vacua of the dual gauge theory can be obtained by
imposing the corresponding $D$- and $F$-flatness conditions that
follow from the lagrangian of the theory. Let us recall that, apart
from the $N_f$ chiral (antichiral) fundamentals ${\cal Q}^i$
($\tilde{\cal Q}_i$), the theory has three adjoint hypermultiplets
$\Phi_I$ ($I=1,2,3$).  We will denote the (scalar) bottom components
of the superfields by lowercase letters.  The $\phi_I$ are the
three complex adjoint scalars of ${\cal N}=4$ SYM and thus they
correspond to the transverse scalar fluctuations of the D3-brane. In
particular, $\phi_3$ represents the fluctuations of the D3 along the
plane orthogonal to the flavor D7-brane. Therefore $\Phi_3$ is
coupled to the fundamentals through a term $\tilde{\cal
Q}_i(m+\Phi_3){\cal Q}^i$, where $m$ is the mass of the fundamentals,
which is proportional to the separation between the D3- and D7-branes.

The $F$-terms corresponding to the fundamental hypermultiplets, vanish
by imposing:
\bear
&&\phi_3={\rm diag}\left(\tilde m_1,\dots, \tilde m_{N-k}, -m,\dots,
-m\right),\label{phi3vev}\\
&&q^i=\left(0\dots 0,q_1^i\dots,q_{k}^i\right)^T,
\quad\tilde q_i=\left( 0\dots 0,\tilde{q}^1_i\dots,
\tilde{q}^{k}_{i}\right),\label{qvev}
\eear
where the number of $m$s in (\ref{phi3vev}) is $k$ and thus in order
to have $\phi_3$ in the Lie algebra of $SU(N)$,  one must have
$\Sigma_{j=1}^{N-k}\tilde{m}_j=k\,m$.
As the quark VEV in this solution has some components which are zero
and others that are different from zero, this choice of vacuum leads to a
mixed Coulomb-Higgs phase. Furthermore, the vanishing of the $F$-terms
associated to the adjoint hypermultiplets gives rise to:
\bear
&&[\phi_1,\phi_3]=[\phi_2,\phi_3]=0,\label{phi3comm}\\
&&q^i\tilde{q}_i+[\phi_1,\phi_2]=0,\label{ftermadj}
\eear
and in view of the vacuum election (\ref{qvev}), the matrices
$\phi_1$ and $\phi_2$ in (\ref{ftermadj}) can be considered as
$k\times k$ matrices given by the lower $k\times k$ block of the
original $N\times N$ matrices. Eq. (\ref{ftermadj}) implies that a
non-vanishing VEV of the quark fields $q$ and $\tilde q$ induces a
non-zero commutator of the adjoint fields $\phi_1$ and $\phi_2$.
These scalars correspond to the four directions transverse to the D3
which lie on the worldvolume of the D7-brane. The dual description in
terms of D7-brane probes must therefore involve a non-trivial
configuration of the worldvolume gauge field along those directions.  
Finally, the $D$-flatness condition results in:
\begin{equation}
|q^i|^2-|\tilde{q}_i|^2+[\phi_1,\phi_1^{\dagger}]+[\phi_2,
\phi_2^{\dagger}]=0\,,\label{dterm}
\end{equation}
where again $\phi_1$ and $\phi_2$ are $k\times k$ matrices.
The constraints (\ref{phi3comm}), (\ref{ftermadj}) and (\ref{dterm})
define the mixed Coulomb-Higgs phase of the theory.


\subsection{The gravity dual}
\label{gravdual}
There is a one to one correspondence between the Higgs phase of
${\cal N}=2$ gauge theories and the moduli space of
instantons \cite{MRD1, MRD2,W}. This follows from the fact that the $F$-
and $D$-flatness conditions can be mapped to the ADHM equations (see
\cite{tong}  for a review). Consequently, one can enter the Higgs phase of
the gauge theory of the D3-D7 system by adding instantonic DBI flux in
the subspace transverse to the D3s (where the gauge theory
lives) but contained in the D7-brane, thus naturally realizing  the
Higgs phase-ADHM equations map.

As we have already said, the gravity dual of the D3-D7 system
consists of $N_f$ D7-brane probes in the $AdS_5\times S^5$ geometry
sourced by $N$ D3-branes. The metric can be written as:
\begin{equation}
ds^2=\frac{\rho^2+\vec{z}^{\,2}}{R^2}\,dx^2_{1,3}
+\frac{R^2}{\rho^2+\vec{z}^{\,2}}(d\vec{y}^{\,2}+d\vec{z}^{\,2}),
\label{ads5metric}
\end{equation}
with $R^4=4\pi\, g_s\,N(\alpha')^2$. The D3-branes lie along the
$\RR^{1,3}$ given by $dx^2_{1,3}$. Notice 
that the six-dimensional subspace orthogonal to the D3-branes is
spanned by the coordinates $\vec y=(y^1,\dots,y^4)$ and $\vec
z=(z^1,z^2)$, with $\rho^2=\vec y\cdot\vec y$ and $r^2=\rho^2+\vec
z^{\,2}$. Additionally, the SUGRA solution includes a four-form RR
potential:
\begin{equation}
C^{(4)}=\left(\frac{r^2}{R^2}\right)^2\ dx^0\wedge\cdots\wedge
dx^3.\label{C4}
\end{equation}

We shall now embed a stack of $N_f$ D7-brane probes in this geometry.
Let us take $\xi^a=(x^{\mu},y^i)$ as worldvolume coordinates
and consider an embedding in which $|\vec{z}|=L$. $L$ represents the
constant transverse separation between the two stacks of D3- and
D7-branes and as a consequence, the D3-D7 strings have a mass
$L/2\pi\alpha'$ which corresponds to the quark mass in the field
theory dual. In addition, let us assume that the worldvolume field  
strength $F$ has non-zero entries only along the subspace
parametrized by the $y^i$ coordinates: $F_{y^iy^j}\equiv F_{ij}$.  It
is straightforward to see that for this configurations the DBI and WZ
terms of the worldvolume action of the D7-branes take the form:
\bear
&&S_{DBI}^{D7}=-T_{7}\int d^4x\,d^4y\,
{\rm Str}\,\sqrt{1+{1\over 2}\left(\frac{\rho^2+L^2}{R^2}\right)^2 F^2
+\frac{1}{16}\left(\frac{\rho^2+L^2}{R^2}\right)^{4}
\left({}^*FF\right)^2}\,,\label{DBId3d7}\\
&&S_{WZ}^{D7}=T_{7}\int d^4x\,d^4y\;
{\rm Str}\left[\frac{1}{4}\left(\frac{\rho^2+L^2}{R^2}\right)^{2}
{}^*FF\,\right],\label{WZd3d7}
\eear
where ${}^*FF={}^*F_{ij}F_{ij}$ and
${}^*F_{ij}=1/2\,\epsilon_{ijkl}\,F_{kl}$. Though the entries in
(\ref{DBId3d7}) and (\ref{WZd3d7}) are $SU(N_f)$ matrices, inside the
symmetrized trace (Str) they can be considered as commutative numbers.

We shall now consider a configuration in which $F$ is selfdual along
the internal $\RR^4$ spanned by the $y^i$ coordinates. In
this case one has ${}^*F=F$ and (\ref{DBId3d7}) becomes the square
root of a perfect square. Consequently, the full action for a
self-dual configuration is just:
\begin{equation}
S^{D7}({\rm selfdual})=-T_7\int d^4x\,d^4y\,
{\rm Str}\, \big[1\big]=-T_7\,N_f\int d^4x\,d^4y.
\label{completeactionD7}
\end{equation}
Notice that the total action does not depend on $L$. This no-force
condition is an explicit manifestation of the SUSY of the system.

The role played by the instantonic worldvolume gauge field can be
better understood by recalling that for selfdual configurations the
integral of the Pontryagin density ${\cal P}(y)$ is quantized for
topological reasons:
\begin{equation}
\int d^4y\,{\cal P}(y)=k\,,\qquad {\rm with}\quad
{\cal P}(y)\equiv\frac{1}{16\pi^2}{1\over (2\pi\alpha')^2}\,
{\rm tr}\,\left[{}^*FF\right],
\label{instquant}
\end{equation}
where $k\in{\mathbb Z}$ is the instanton number.
Then, by looking at the WZ term of the action one can check that a
worldvolume gauge field satisfying (\ref{instquant}) has the effect
of inducing $k$ units of D3-brane charge in the D7-brane
worldvolume. Finally, it is worth to remark that the existence of
these instantonic configurations is possible because we are
considering $N_f>1$ flavor D7-branes thus having a non-abelian
worldvolume gauge theory.


\subsection{Microscopic description}
\label{microdescrip}

In this subsection we will describe the flavor D7-branes with $k$
D3-branes dissolved on them in terms of $k$ dielectric D3-branes
\cite{M} expanding to a transverse fuzzy $\RR^4$ and thus
reconstructing the flavor D7-brane. In this microscopic description
one has a stack of $k$ D3-branes in the background (\ref{ads5metric}).
These D3-branes are extended along the four Minkowski coordinates
$x^{\mu}$, whereas the transverse coordinates $\vec{y}$ and $\vec{z}$
must be regarded as the matrix scalar fields $Y^i$ and $Z^j$,  taking
values in the adjoint representation of  $SU(k)$. Since we are
interested in a setup where the D3-branes are localized in the
subspace transverse to the would-be D7-branes, we must take the
scalars $Z^j$ to be abelian. The dynamics of the dielectric
D3-branes is determined by the Myers dielectric action \cite{M} which
is the sum of a DBI and WZ term. For our configuration the DBI term 
reads:
\begin{equation}
S_{DBI}^{D3}=-T_3\int d^4x\,{\bf Str}\left[\left({\hat r^{\,2}
\over R^2}\right)^2\sqrt{\det\left(\delta_{ij}-{R^2\over \hat r^{2}}
\theta_{ij}\right)}\,\right],
\label{dieldbi}
\end{equation}
where $\hat r^2$ is the matrix $\hat r^{\,2}=(Y^i)^2+Z^2$, ${\bf
Str}$ represents the symmetrized trace over the $SU(k)$ indices and
we have defined the matrix $\theta_{ij}$ as:
\begin{equation}
i\theta_{ij}\equiv [Y^i,Y^j]/(2\pi\alpha').
\label{thetaij}
\end{equation}
Hence $\theta_{ij}$ is antisymmetric  in the $i,j$ indices and, as a
matrix of $SU(k)$, is hermitian. Reading $C^{(4)}$ from (\ref{C4}),
the WZ term of the action can be written as:
\begin{equation}
S_{WZ}^{D3}=T_{3}\int d^4x\,{\bf Str}\left[\left({\hat r^{\,2}\over
R^2}\right)^2\right].
\label{dielwz}
\end{equation}
We shall now assume that $\theta_{ij}$ is selfdual with respect to
the $ij$ indices, \ie \
$1/2\,\epsilon_{ijkl}\,\theta_{kl}=\theta_{ij}$, so that there are
only three independent matrices: $\theta_{12}=\theta_{34}$,
$\theta_{13}=\theta_{42}$ and $\theta_{14}=\theta_{23}$. Then, it is
not difficult to see that the integrand of (\ref{dieldbi}) becomes
the square root of a perfect square, and the action of the system,
given by the sum of (\ref{dieldbi}) and (\ref{dielwz}), reduces to:
\begin{equation}
S^{D3}({\rm selfdual})=-{T_3\over 4}\int d^{4}x\,{\bf Str}\left[
\theta^2\right]=-\pi^2\,T_7(2\pi\alpha')^2
\int d^{4}x\,{\bf Str}\left[\theta^2\right],
\label{completeactionD3}
\end{equation}
and we have rewritten the result in terms of the tension of the
D7-brane since it will be useful below.

This microscopic description in terms of color D3-branes should match
the field theory analysis performed at the beginning of section
\ref{d3d7inters}. In particular, we expect that the $F$- and
$D$-flatness conditions of the adjoint scalars are the same as the
ones satisfied by the transverse scalars of the dielectric
D3-branes. Let us then define the following complex combinations of
the $Y^i$ matrices:
\begin{equation}
2\pi\alpha'\,\phi_1\equiv (Y^1+iY^2)/\sqrt{2}\,,
\qquad\quad
2\pi\alpha'\,\phi_2\equiv (Y^3+iY^4)/\sqrt{2}\,,
\label{dielphi12}
\end{equation}
where we have introduced the factor $2\pi\alpha'$ to take into
account the standard relation between coordinates and scalar fields
in string theory. With these identifications one can easily compute
the commutators appearing in the equations defining the mixed
Coulomb-Higgs phase. Actually, the eqs. (\ref{ftermadj}) and
(\ref{dterm}) become:
\begin{equation}
q^i\tilde{q}_i=(\theta_{23}-i\theta_{13})/(2\pi\alpha')\,,\qquad\qquad
|\tilde{q}_i|^2-|q^i|^2=\theta_{12}/(\pi\alpha')\,,
\label{fdtermsmicro}
\end{equation}
where we have already taken into account the selfduality of
$\theta_{ij}$. Furthermore, the adjoint scalar $\phi_3$ is
proportional to $Z^1+iZ^2$ and remember that $Z^1$ and $Z^2$ are
abelian, reflecting the fact that the D3-branes are localized  in
these directions. Hence one has $[\phi_1,\phi_3]=[\phi_2,\phi_3]=0$
and the constraint (\ref{phi3comm}) is then fulfilled. Notice that
eq. (\ref{fdtermsmicro}) provides an identification between the
$\theta$s and the VEVs of the fundamentals. Remarkably,
(\ref{fdtermsmicro}) is telling us that having a non-zero VEV of the
fundamentals is only possible if the transverse scalars
$Y^i$ of the D3-branes have non-zero commutators, and thus the
D3-branes get polarized along these directions.  

It would now be interesting to relate the present microscopic
description in terms of dielectric color D3-branes to the macroscopic
description of subsection \ref{gravdual} in terms of flavor D7-branes. To
this aim, let us compare the WZ terms of the D7- and D3-branes given
by (\ref{WZd3d7}) and (\ref{dielwz}). This motivates a map between
$k\times k$ matrices in the microscopic description and functions of
the $y$ coordinates in the macroscopic approach. Indeed, a matrix
$\hat f$ is mapped to the function $f(y)$ according to the following
rule:
\begin{equation}
{\bf Str} \,[\,\hat f\,]\,\Longrightarrow\, \int d^4y\, {\cal
P}(y)\, f(y),
\label{micro-macro}
\end{equation}
where the kernel ${\cal P}(y)$ on the right-hand side of
(\ref{micro-macro}) is the Pontryagin density defined in
(\ref{instquant}). Actually, the comparison between both WZ actions
tells us that the matrix $\hat r^2$ is mapped to the function $\vec
y^{\,2}+\vec z^{\,2}$. Notice also that when
$\hat f$ is the unit
$k\times k$ matrix and $f(y)=1$ both sides of (\ref{micro-macro}) are 
equal to the instanton number $k$ (see eq. (\ref{instquant})).
Furthermore, the comparison of the complete actions of the D7- and
D3-branes, given by (\ref{completeactionD7}) and
(\ref{completeactionD3}) respectively, leads to:
\begin{equation}
(2\pi\alpha')^2\,{\bf Str} [\,\theta^2\,]\,
\Longrightarrow\, \int d^4y\,{N_f\over \pi^2}.
\label{theta-map}
\end{equation}
Thus, according to the general rule (\ref{micro-macro}), the function
corresponding to $\theta^2$ is given by:
\begin{equation}
(\,2\pi\alpha'\,)^2\,\theta^2\,\,
\Longrightarrow \,\,{N_f\over \pi^2 \,\,{\cal P}(y)}\,\,.
\label{theta-instanton}
\end{equation}
This relation confirms the realization of the Higgs phase of the dual
theory through the addition of instantonic flux of the worldvolume
gauge field to the flavor D7-branes. Notice that $\theta^2$ is a
measure of the non-commutativity of the adjoint scalars and
(\ref{theta-instanton}) implies that this non-commutativity is
related to the (inverse of the) Pontryagin density of the macroscopic
D7-branes. In addition, $\theta$ is related to the VEVs of the
fundamentals through (\ref{fdtermsmicro}). Therefore, eq.
(\ref{theta-instanton}) implies a relation between the quark VEV and
the instanton density on the flavor D7-branes. Let us make this
correspondence more precise by studying the one-instanton
configuration of the $N_f=2$ gauge theory on the D7-branes. In the
so-called singular gauge the $SU(2)$ gauge field is given by:
\begin{equation}
{A_i\over 2\pi\alpha'}=2i\Lambda^2\frac{\bar{\sigma}_{ij}\,y^j}
{\rho^2(\rho^2+\Lambda^2)}\,,\qquad{\rm with}\quad
\bar{\sigma}_{ij}={1\over 4}\left(\bar\sigma_i\,\sigma_j-
\bar\sigma_j\,\sigma_i\right),\quad
\sigma_i=(i\vec \tau,1_{2\times 2}),
\label{su2field}
\end{equation}
where $\vec\tau$ are the Pauli matrices (and
$\bar\sigma_i=\sigma_i^\dagger$), $\rho^2=\vec y\cdot\vec y$ and
$\Lambda$ is a constant (the instanton size). Computing $F_{ij}$ and
inserting it into the expression for ${\cal P}(y)$ given in
(\ref{instquant}) one arrives at:
\begin{equation}
{\cal P}(y)={6\over \pi^2}\,{\Lambda^4\over
(\rho^2+\Lambda^2)^4}\,.
\label{su2pontr}
\end{equation}
It can be easily checked that (\ref{su2pontr}) satisfies
(\ref{instquant}) with $k=1$.
This result can shed light on the relation between the Higgs
mechanism of the field theory and the instanton density in the
holographic description. For simplicity we will assume that all quark
VEVs are proportional to some scale $v$: $q,\, \tilde q\sim v$. Hence
(\ref{fdtermsmicro}) implies $\theta\sim\alpha'\,v^2$, and
plugging this last result and (\ref{su2pontr}) into
(\ref{theta-instanton}) one gets the following relation:
\begin{equation}
v\sim{\rho^2+\Lambda^2\over\alpha'\Lambda}\,,
\label{holoVEV}
\end{equation}
where $\rho$ should be regarded as the energy scale of the gauge
theory. In the far IR ($\rho\approx0$) (\ref{holoVEV}) reduces to:
\begin{equation}
v\sim\Lambda/\alpha',
\label{v-Lambda}
\end{equation}
which, up to numerical factors, is precisely the relation between the
quark VEV and the instanton size that has been  obtained in
\cite{EGG}. As for the full expression (\ref{holoVEV}), notice that
for any finite non-zero $\rho$ the quark VEV never vanishes and goes to
infinity in both the large and small instanton limits. However, in
the far IR the quark VEV goes to zero in the small instanton limit.
This region should be clearly singular because a zero quark VEV would
correspond to an {\it unhiggsed} theory and thus extra light degrees
of freedom must appear.


\subsection{Fluctuations}

Finally, we will determine the meson spectrum by studying the
fluctuations around the instanton configuration described in subsection
\ref{gravdual}.
Actually, we will focus on the fluctuations of the
worldvolume gauge field $A=A^{inst}+a$. Then, the total field strength
on the D7-brane worldvolume is $F_{ab}=F_{ab}^{\rm inst}+f_{ab}\,$ ($a,b$
run over the worldvolume), $F_{ab}^{\rm inst}$ is the selfdual
field strength of
$A^{inst}$ and
$f_{ab}$ is given by:
\begin{equation}
f_{ab}=\partial_{a}a_{b}-\partial_{b}a_{a}+[A^{inst}_{a},a_{b}]/
(2\pi\alpha')+[a_{a},A^{inst}_{b}]/(2\pi\alpha')+[a_{a},a_{b}]/
(2\pi\alpha').
\label{fab}
\end{equation}
We shall consider the DBI and WZ terms of the action for a stack of
D7-branes with worldvolume coordinates $\xi^a=(x^{\mu},y^i)$, an
embedding $|\vec z|=L$ (as in subsection \ref{gravdual}) and with the
worldvolume gauge field strength $F_{ab}$ just described.
The mass spectrum will be determined by the equation of motion resulting
from the action expanded up to quadratic order in the fluctuations.
 
\begin{figure}[]
\centerline{\includegraphics{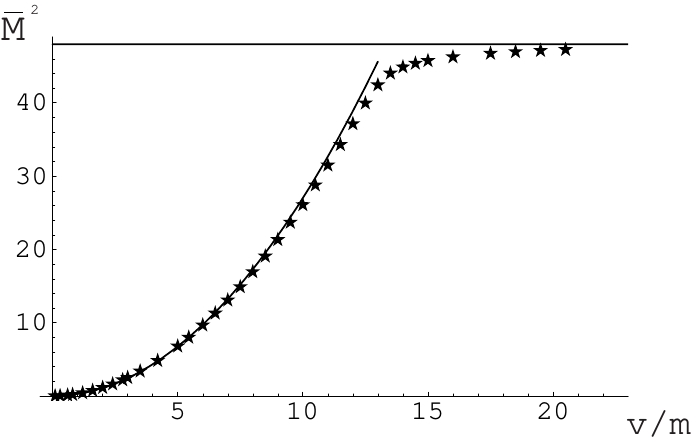}}
\caption{Numerical values of $\bar M^2=R^4L^{-2}M^2$ as a function of the
ratio of the quark VEV $v=\Lambda/(2\pi\alpha')$ and the quark mass
$m_q=L/(2\pi\alpha')$. The solid line represents the prediction obtained
by the WKB method. For large $v$, $\bar M$ becomes independent of both
$m_q$ and $v$ as in \cite{EGG}, while the mass gap vanishes in the small
instanton limit $v\to0$.}
\label{spc}
\end{figure}
 
Let us concentrate on the $N_f=2$ one-instanton configuration described
above and restrict ourselves to fluctuations with $a_i=0$, \ie \ those
with non-zero components only along the Minkowski directions.
Additionally, we assume that
$a_{\mu}^{(l)}=\xi_{\mu}(k)\,f(\rho)\,e^{ik_{\mu}x^{\mu}}\,\tau^l$, with
$k^\mu\xi_\mu=0$. This ansatz automatically solves the equation of motion
of $a_i$. The equation for $a_\mu$ reduces to a second order differential
equation for $f(\rho)$. This equation depends on $M^2=-k^2$ and has
normalizable solutions for a discrete set of values of $M$ giving rise to
a discrete mass spectrum. We compute this spectrum by applying a shooting
technique, obtaining the result plotted in fig \ref{spc}.

\section*{Acknowledgments}

I would like to thank the organizers of the 3rd RTN
Workshop (Valencia, October 2007) where this work was presented.  The
results presented here were obtained in collaboration with Alfonso V.
Ramallo and Diego Rodr\'\i guez G\'omez. I am grateful to Alfonso V.
Ramallo and Jonathan P. Shock for useful discussions. This  work was
supported in part by MEC and  FEDER  under grant FPA2005-00188,  by the
Spanish Consolider-Ingenio 2010 Programme CPAN (CSD2007-00042), by Xunta
de Galicia (Conseller\'\i a de Educaci\'on and grant PGIDIT06PXIB206185PR)
and by  the EC Commission under  grant MRTN-CT-2004-005104.

\end{document}